\documentclass{ws-mpla}

\newcommand{\be}{\begin{equation}}
\newcommand{\ee}{\end{equation}}
\newcommand{\bea}{\begin{eqnarray}}
\newcommand{\eea}{\end{eqnarray}}

\begin{document}

\markboth{Massimiliano Rinaldi}
{Particle production and transplanckian problem on the non-commutative plane}

\catchline{}{}{}{}{}

\title {\bf PARTICLE PRODUCTION AND TRANSPLANCKIAN PROBLEM ON THE NON-COMMUTATIVE PLANE}

\author{\footnotesize MASSIMILIANO RINALDI}

\address{D\'epartment de Physique 
Th\'eorique, Universit\'e de Gen\`eve, \\
24 quai Ernest Ansermet  CH--1211 Gen\`eve 4, Switzerland\\ massimiliano.rinaldi@unige.ch}

\maketitle

\begin{abstract}

\noindent We consider the coherent state approach to non-commutativity, and we derive from it an effective quantum scalar field theory. We show how the non-commutativity can be taken in account by a suitable modification of the Klein-Gordon product, and of the equal-time commutation relations. We prove that, in curved space, the Bogolubov coefficients are unchanged, hence the number density of the produced particle is the same as for the commutative case. What changes though is the associated energy density, and this offers a simple solution to the transplanckian problem. 

\end{abstract}

\ccode{04.60.Bc, 04.62.+v}

\section{Introduction}

\noindent In recent years, we have seen many proposals aimed to quantize consistently the gravitational field from fundamental principles. From a phenomenological point of view, a more modest approach consists in introducing reasonable modifications to quantum field theory and look for observable consequences in black holes or inflationary models. For example, one can construct a theory where the dispersion relations depart from linearity above a certain energy scale, thus breaking local Lorentz invariance. This approach is motivated by analogue models of gravity in condensed matter systems \cite{NuovoCimento}, by deformations of the Lorentz group \cite{AmelinoCamelia}, or by tensor-vector models of gravity \cite{Jacob}. Modified dispersion relations were considered in the context of renormalization,  and particle production in curved spacetime \cite{MaxMazz,max1,MaxMazz2,max2}. In the latter case, the common result is that the thermal spectrum seen from an accelerated detector or from an asymptotic observer on a black hole background is only marginally affected by non-linear dispersion relations \cite{UnruhOrig,UnruhShuz,UnruhMax}. 

There exists an alternative proposal, based on a new symmetry of the path integral duality \cite{paddy}. In this case, the modification directly brings a minimal length in the propagator, which becomes finite in the ultraviolet regime. Starting from different hypothesis, the same propagator was reconsidered also in other works \cite{pepe}. In both cases, the form of the field modes, associated to the modified propagator is unknown, therefore it is difficult to evaluate exactly some effects. 

In contrast to these proposals, here we would like to study how quantum field theory is modified when spacetime has an intrinsic non-commutative structure. This topic has been intensively investigated by assuming that non-commutative effects in field theory are implemented by replacing the ordinary product among functions with the so-called star-product \cite{NCclassic,NCclassic2,starproduct,starproduct2,starproduct3}. 
Instead, in this paper we would like to take onboard an alternative point of view, and consider the coherent state approach to non-commutativity introduced in a series of papers by E.\ Spallucci and collaborators \cite{Sma1,Sma2,Sma3,Sma4}. As we will briefly explain below, this model does not need the star-product, since all non-commutative effects are encoded in the Gaussian damping of the field modes \footnote{On mathematical grounds, in this theory the star-product can be seen as replaced by the so-called Voros product \cite{voros,voros2}.}. As a result, the field propagator is finite in the ultraviolet limit, but the dispersion relation is the same as the relativistic one.  Compared to the wider class of modified theories mentioned above, this proposal has a stronger predictive power, as both field modes and propagators are known.
For example, this model has been already studies in connection with  the Casimir effect \cite{Casadio}, and inspired several works on black holes \cite{Nico,Nico2,Nico3,Nico4,Nico5,Nico6,Nico7,Nico8,Nico9,Nico10}, Unruh effect \cite{Unruh}, inflation \cite{MaxInf}, and quantum gravity \cite{NicoSp}. 

The plan of the paper is the following: in the next section, we recall the main features of the coherent state approach to non-commutativity, and we construct a massive scalar field living on a two-dimensional Minkowski plane. In section III, we consider the generalization in curved space, and we show that Bogolubov transformations are not affected in the fourth section. In section IV, we look at the transplanckian problem of a black hole and compare with analogous calculations for the Unruh effect. Finally, we conclude the paper with few remarks.  

\section{Non-commutative field theory}

\noindent To begin with,  let $\hat z_1, \hat z_2$ be the coordinate operators of a two-dimensional non-commutative plane, that satisfy the algebra \cite{Sma1}
\bea
[\hat z_1, \hat z_2]=i\theta\ .
\eea
Generalizations to higher even-dimensional spaces are straightforward \cite{Sma2}. One can define the new operators
\bea
\hat A=\hat z_1+i\hat z_2\ ,\quad \hat A^{\dagger}=\hat z_1-i\hat z_2\ ,
\eea
which satisfy the canonical commutator $[\hat A,\hat A^{\dagger}]=2\theta$. The coherent states associated to these operators are the states $|\alpha\rangle$ such that $
\hat A|\alpha\rangle=\alpha  |\alpha\rangle$.  Their explicit form reads
\bea
|\alpha\rangle=\exp\left[ {1\over \theta}\left(\alpha^*\hat A-\alpha\hat A^{\dagger}\right)\right]|0\rangle\ ,
\eea
and one can show that $\langle\alpha|\alpha\rangle=1$. Physical, commuting coordinates are c-numbers, defined as the expectation values on coherent states of the coordinate operators, namely
\bea
x_1\equiv\langle \alpha | \hat z_1 |\alpha\rangle={\rm Re}(\alpha)\ ,\quad x_2\equiv\langle \alpha | \hat z_2 |\alpha\rangle={\rm Im}(\alpha)\ .
\eea
The vector $(x_1,x_2)$ is interpreted as the mean position of the point particle on the non-commutative plane. So, how does non-commutativity show up in quantum field theory? The answer lies essentially in the fact that, now, every function $F(\vec x)$ must be promoted to an operator $\hat F(\hat z_1,\hat z_2)$ evaluated on coherent states. In particular, this holds for the monochromatic wave function, which is turned into  the operator $\exp(i\vec p\cdot\hat{\vec z}\,)$, where $\vec p=(p_1,p_2)$ is a real vector. After having defined the transverse momenta $p_{\pm}=(p_1\pm ip_2)/2$, one can utilize the Baker-Campbell-Hausdorff formula to find \footnote{We stress that the components of $\vec p$ are c-numbers, so they act trivially on the coherent states. There is a more formal approach in terms of non-commutative Fock space, but the results are the same \cite{voros}. }
\bea\label{HC}
\langle \alpha|  e^{ip_1\hat z_1+ip_2\hat z_2  } |\alpha\rangle&=&\langle \alpha|  e^{{i\over 2}p_+\hat A^{\dagger}} e^{{i\over 2}p_-\hat A} |\alpha\rangle e^{ {p_+p_-\over 4}[A^{\dagger},A]} =e^{i\vec p\cdot\vec x-{\theta\over 4}\,(p_1^2+p_2^2)}\ .
\eea
We see that the main effect of the intrinsic non-commutative structure of space resides in the damping term embedded in the plane wave operator. We will show later that this term can also be interpreted as a deformation of the measure in the Fourier integral. There is one crucial aspect in the expression (\ref{HC}): the relative sign in the quadratic term $(p_1^2+p_2^2)$ is \emph{independent} of the signature of the metric on the plane. It simply arises from the product $p_-p_+$, hence its form is the same in both Euclidean and Lorentzian signatures.

Now, let us consider a scalar field $\phi(t,x)$ with mass $m$, which propagates on a two-dimensional  Minkowski space, and is governed by the standard Klein-Gordon equation \footnote{In our notation, $ds^2=dt^2-dx^2$. }
\bea\label{KGeq}
(\square+m^2)\phi(t,x)=0
\eea
According to the prescriptions above, the positive frequency field modes are modified according to
\bea\label{modes}
u_p(t,x)={e^{-\ell^2(\omega^2+p^2)}\over \sqrt{4\pi \omega}}e^{-i\omega t+i\vec p\cdot \vec x} \ .
\eea
In this expression, we identify $p_1$ with $\omega=\sqrt{p^2+m^2}$, $\ell^2=\theta/4$, and we set $p_2=p$. As the modes are solutions to Eq.\ (\ref{KGeq}), we define the corresponding Klein-Gordon product \cite{BirDav}, which, however, must be modified to accommodate to the different normalization of the modes. Thus, we have a ``damped'' $\delta$-function
\bea\label{KGprod}
(u_p,u_{p'})\!=\!-i\int dx (u_p \overleftrightarrow{\partial_t} u_{p'}^*)=e^{-2\ell^2(\omega^2+p^2)}\delta(p-p'),
\eea
and this reflects the fact that coherent states are normalized but not orthogonal. The scalar field can be represented as the usual mode sum
\bea
\phi(t,x)=\int {d\vec p \over \sqrt{4\pi\omega}}\left[\hat a_pu_p(t,x)+\hat a_p^{\dagger}u_p^*(t,x)  \right] \ ,
\eea
where $\hat a_p$ is the annihilation operator, which fulfills the standard rule $[\hat a_p,\hat a_{p'}^{\dagger}]=4\pi\omega\delta(p-p')$. It follows that the equal-time commutator reads
\bea
[\phi(t,x),\dot\phi(t,x')]\!=\!{i\over 4\sqrt{\pi}\ell}\,e^{-2\ell^2 m^2-(x-x')^2/ 16\ell^2}\ .
\eea
In the vanishing $\ell$ limit, the  right-hand side smoothly tends to $i\delta(x-x')$. The Wightman's positive frequency function can be determined in the usual way, and the result is
\bea
G^+(x^{\mu},x'^{\mu})\equiv\langle 0|\phi(x^{\mu})\phi(x'^{\mu})|0\rangle=\int  {d\vec p\over 4\pi\omega}\,e^{-2\ell^2(\omega^2+p^2)-ip_{\mu}(x^{\mu}-x'^{\mu})}\ ,
\eea
from which it follows that the Feynman propagator reads
\bea
G_F(x^{\mu},x'^{\mu})=-i\int {d\vec p\over 4\pi \omega}\,e^{-2\ell^2(\omega^2+p^2)} \left[\theta(t\!-\!t')e^{-ip_{\mu}(x^{\mu}-x'^{\mu})}\!+\theta(t'\!-t)e^{ip_{\mu}(x^{\mu}-x'^{\mu})}\right]\!.
\eea
This expression can be derived from the integral (for $p_1=\omega$ and $p_2=p$)
\bea\label{momprop}
G_F(x^{\mu},x'^{\mu})=i\int {d^2p\over (2\pi)^2}{\,e^{-2\ell^2(p_1^2+p_2^2)-ip_{\mu}(x^{\mu}-x'^{\mu})}\over p^2_1-p_2^2-m^2}\ ,
\eea
from which we can easily read off the momentum space propagator
\bea\label{Fprop}
\tilde G_F(p_1,p_2)={e^{-2\ell^2(p_1^2+p_2^2)}\over p^2_1-p_2^2-m^2}\ ,
\eea
which is the same found   via path integral quantization \cite{Sma1}. It is easy to check that it satisfies the equation
\bea
(\square+m^2)G_F(x^{\mu},x'^{\mu})=-{i\over 8\pi \ell^2}e^{-\left(\Delta t^2+\Delta x^2\right)/8\ell^2} ,
\eea
where $\Delta t^2=(t-t')^2$ and $\Delta x^2=(x-x')^2$. The right-hand side becomes the usual $\delta$-function for $\ell\rightarrow 0$.

The above integrals are not trivial, and it is difficult to find an expression for the Lorentzian propagator in coordinate space. However, it is easy to find it in Euclidean space, and for many applications this is sufficient to obtain valuable results. The Euclidean version of (\ref{Fprop}) that we would have obtained by working on a Euclidean non-commutative plane, can be written in the Schwinger formalism as
\bea
\tilde G_F^E(p_1,p_2)={e^{-2\ell^2(p_1^2+p_2^2)}\over p^2_1+p_2^2+m^2}=e^{\ell^2m^2}\int_{\ell^2}^{\infty}ds\,e^{-s(p_1^2+p_2^2+m^2)}\ .
\eea
The corresponding Euclidean propagator in coordinate space is  
\bea
G^E_F(x^{\mu},x'^{\mu})=e^{\ell^2m^2}\int_{\ell^2}^{\infty}ds\int_{0}^{\infty}{dp_1dp_2\over 2\pi^2}\,e^{-ip_1x_1-ip_2x_2-s(p_1^2+p_1^2+m^2)}\ .
\eea
By integrating, we find the following  cases:
\begin{itemize}
\item for $\ell = 0$ we have the usual massive two-dimensional scalar propagator, which shows a logarithmic divergence as $m\rightarrow 0$ or $(\Delta x_1^2+\Delta x_2^2)\rightarrow 0$; 
\item for $\ell\neq 0$ and $m=0$, the integral does not converge for $s\rightarrow \infty$. However, one can integrate between $\ell^2$ and some arbitrary $L^2>\ell^2$, and then expand for $(\Delta x_1^2+\Delta x_2^2)\rightarrow 0$. The leading term of the series is constant and proportional to $\ln (K/L)$, thus it is irrelevant when quantities obtained by differentiating the propagator, such as the stress tensor, are computed. We note that this situation is peculiar to the two-dimensional case only, as the integral contains the factor $1/s$. In four dimensions, this factor is $1/s^2$ so the integral converges for $s\rightarrow \infty$ \cite{Unruh};
\item in the case $\ell\neq 0$ and $m\neq 0$, the integral cannot be computed analytically. However, a simple numerical computation shows that the propagator is finite in the coincident limit for any non-zero $\ell$, and this confirms the ultraviolet regulating nature of the minimal length induced in the field theory by non-commutativity.
\end{itemize}
Let us now look at the energy density. The Hamiltonian operator for the scalar field becomes
\bea\label{Ham}
\hat H={1\over 2}\int d^2x\left[\dot \phi^2+(\vec \nabla \phi)^2+m^2\phi^2\right]={1\over 2}\int d\vec p \,e^{-2\ell^2(p^2+\omega^2)}\omega\left(\hat{a}_p\hat{a}_p^{\dagger}+\hat{a}_p^{\dagger}\hat{a}_p\right)\ .
\eea
This expression clearly renders unnecessary the normal ordering, usually employed to eliminate the divergent zero-point energy. In fact
\bea
\langle 0| \hat H|0\rangle=e^{-2\ell^2 m^2}\int_0^{\infty}dp\,e^{-4\ell^2 p^2}\sqrt{p^2+m^2}\ ,
\eea
which can be shown numerically to be convergent for any $m> 0$. In the massless case, the above integral gives simply $(8\ell^2)^{-2}$.

\section{Particle production}

\noindent We now turn to the problem of the quantization of the field on a curved space. It is well known that, in a general space-time, one can find more than one complete orthonormal sets of modes, and construct the fields as different mode sums. The transformation between two inequivalent basis generates non trivial Bogolubov coefficients, which can be interpreted as a sign of particle production \cite{BirDav}. Let us assume that the Klein-Gordon product (\ref{KGprod}) can be locally carried over curved space, and let $u_i$ and $v_i$ be two basis of normal modes of the type (\ref{modes}). The scalar field can be represented as
\bea
\phi(x)=\displaystyle\sum_i(\,\hat a_iu_i+{\rm h.c.})=\displaystyle\sum_j(\,\hat b_jv_j+{\rm h.c.})\ .
\eea
Here, $\hat a$ and $\hat b$ are the inequivalent annihilation operators, which are assumed to satisfy the usual commutation rules. We can write the damped modes as $u_i=g_uU_i$ and $v_i=g_v V_i$, where $U_i=u_i(\ell=0)$ and $V_i=v_i(\ell=0)$ are the standard modes of the commutative theory, and $g_{u,v}$ are the Gaussian damping factors. From what we have seen above, the Klein-Gordon product is modified according to $(u_i,u_j)=g_u^2(U_i,U_j)$, together with the analogous expression for $(v_i,v_j)$. Since the modes $U_i$ and $V_j$ are related by the linear transformations \cite{BirDav}
\bea
V_j=\displaystyle\sum_i(\alpha_{ij}U_i+\beta_{ij}U^*_i)\ ,
\eea
it follows that 
\bea
v_j={g_v\over g_u}\displaystyle\sum_i(\alpha_{ij}u_i+\beta_{ij}u^*_i)\ ,\quad u_i={g_u\over g_v}\displaystyle\sum_j(\alpha_{ij}^*v_j-\beta_{ji}v^*_j)\ .
\eea
It is not difficult to prove that 
\bea
\hat b_l= \displaystyle\sum_i(\alpha_{il}^*\hat a_i-\beta_{il}^*\hat a^{\dagger}_i)\ ,
\eea
which means  that the relation between the two sets of annihilation and creation operators are not affected by the Gaussian damping factors. Most importantly, we find that, despite the Klein-Gordon product is modified according to (\ref{KGprod}), the Bogolubov coefficients are unchanged with respect to the commutative case, since
\bea
b_{jl}=-{1\over g_u g_v}(v_j,u^*_l)\equiv (V_j,U^*_l)\ ,
\eea 
and a similar expression holds for $\alpha_{ij}$. From this almost trivial argument, we deduce that the non-commutative structure of spacetime, when treated in terms of coherent states, does not affect the production of particles, given that the Bogolubov coefficients, and in particular $\beta$, are not modified with respect to the commutative case. However, the \emph{energy density} associated to  these particle is very different when compared the commutative case. In fact, even though the expectation value of the number density operator $\langle N_i\rangle=\displaystyle \sum_j |\beta_{ij}|^2$ is unchanged, the energy density associated to it is given by Eq.\ (\ref{Ham}), which, for the $i$-th particle species can be written as
\bea
\hat H_i ={1\over 2}\int d\vec p \,e^{-2\ell^2(p^2+\omega^2)}\omega \left({1\over 2}+\hat N_i\right) \ .
\eea
Therefore, even if there is a conspicuous production of high frequency particles, their contribution to the total energy density is suppressed.

The fact that particle production is not affected by non-commutativity has several positive consequences. In inflationary models, this result suggests that the spectrum of primordial perturbations is unchanged despite the high energy modifications on the Feynman propagator. Moreover, gravitons and particles produced during a transition between two different cosmological eras are indistinguishable from the ones obtained in the commutative case. Thus, a priori, there are no observations that can rule out, so far, this theory. 

\section{Transplanckian problem}

\noindent In the case of  black holes, it is well-known that the Hawking flux measured by an asymptotic observer derives from an accumulation of in-modes of arbitrarily high frequency \cite{NuovoCimento}. This aspect has always created some unease as it entails arbitrarily high energies, which render dubious the validity of the theory itself. However, non-commutative effects prevent this problem, as we quickly show below.

When a free-falling observer near the horizon measures the frequency $\Omega$ of the field modes $\phi\sim (4\pi \Omega)^{-1/2} \exp(i\Omega t-i\Omega x)$  finds
\bea
\Omega\propto \omega \left(1-{2M\over r}\right)^{-1}\ ,
\eea
where $\omega$ is the frequency measured by an asymptotic observer, $M$ is the mass of the hole, and $r$ is the radial coordinate of the Schwarzschild metric, which tend to $2M$ as we approach the horizon at $r_h=2M$ \cite{NuovoCimento}. The free-falling observer can locally treat $\phi$ as a Minkowski field, thus its energy density is the sum of the energy $\hbar \Omega$ of each mode. As this diverges when the observer approaches the horizon, one expect that this huge energy backreacts against the metric, thus invalidating the initial  hypothesis that the field is a test field propagating on a fixed background. However, as we have shown above, when non-commutativity is switched on, each mode is equipped with a damping factor of the form $\exp(-4\ell^2\Omega^2)$ \footnote{For simplicity, we consider massless fields, for which $\omega=p$, as the dispersion relation is not modified.}. Therefore, as the local frequency increases, the mode and its energy is increasingly suppressed, and the energy density of the field is naturally bounded by the scale $1\over \ell$.

These results can explain the apparent contradiction between what was found above and the results  on the Unruh effect \cite{Unruh}. In this work, the authors find that the spectrum of the particles seen by a uniformly accelerated detector is suppressed and no longer thermal, when the propagator is modified according to Eq.\ (\ref{Fprop}). Thus, in contrast with most models with a minimal length, where the Unruh effect is robust \cite{UnruhMax,pepe,UnruhPaddy}, the Gaussian damping in the propagator  has a dramatic impact on the spectrum. This fact can lead to think that if the Unruh effect is calculated via Bogolubov transformations, rather than with an accelerated detector, the result would be the same as for the commutative case, and hence in contradiction with the findings on the Unruh effect. However, in this work, what is found to be suppressed is, in fact, the response rate, which is nothing but an energy flux. Therefore there is no contradiction because here we proved that the energy density of the produced particles is suppressed, and so it must be also its flux through the detector.

\section{Conclusions}

\noindent In this paper we constructed the field theory of a massive scalar field on the non-commutative plane, by mode analysis. The results coincide with the path integral approach, and clarified the relation between the Euclidean and Lorentzian propagator. We then used this field theory to tackle the transplanckian problem for a black hole. The main result is that the fuzziness on the manifold puts an upper limit on the energy density that can be stored near the horizon by a free-falling observer. The quantum backreaction on the geometry of the black hole can in principle be calculated with a suitable effective action, and it represents our next goal. In conclusion, we believe that the coherent state formulation of non-commutativity can offer new and intriguing perspectives on the phenomenology of quantum gravity. In this paper we presented just a glimpse of the potential of this theory, which certainly deserves further investigations.

\section*{Acknowledgments} 

I wish to thank P.\  Nicolini, and E.\ Spallucci for reading the manuscript, and all the people in the  Cosmology Group of the University of Geneva for continuous and  stimulating discussions. This work is supported by the Fond National Suisse.

\end{document}